# Optical identification of atomically thin dichalcogenide crystals

A Castellanos-Gomez[1], N Agraït[1,2,3] and G Rubio-Bollinger[1,2]

[1] Departamento de Física de la Materia Condensada (C–III).
Universidad Autónoma de Madrid, Campus de Cantoblanco, 28049 Madrid, Spain.
[2] Instituto Universitario de Ciencia de Materiales "Nicolás Cabrera".
Universidad Autónoma de Madrid, Campus de Cantoblanco, 28049 Madrid, Spain.
[3] Instituto Madrileño de Estudios Avanzados en Nanociencia
IMDEA-Nanociencia, 28049 Madrid, Spain

Email: gabino.rubio@uam.es

We present a systematic study of the optical contrast of niobium diselenide and molybdenum disulfide flakes deposited onto silicon wafers with a thermally grown silicon oxide layer. We measure the optical contrast of flakes whose thickness, which is obtained by atomic force microscopy AFM, ranges from 200 layers down to a monolayer using different illumination wavelengths in the visible spectrum. The refractive index of these thin crystals has been obtained from the optical contrast using Fresnel law. In this way the optical microscopy data can be quantitatively analyzed to determine the thickness of the flakes in a fast and non-destructive way.

Transition metal dichalcogenides are inorganic layered materials which exhibit a large variety of electronic behaviors such as semiconductivity, superconductivity or charge density wave. Semiconducting dichalcogenides like $MoS_2$ are promising materials for electronic applications. Their mechanical properties make them candidates to be used for flexible field effect transistors[1,2] (FETs), substituting the commonly used organic-based FETs which show a very low charge carrier mobility. Indeed, dichalcogenide FETs present a high mobility[2] comparable to the one of Si FETs. In addition, optical properties of these materials can be exploited for solar cells, photoelectrochemical cells and photocatalytic applications[3,4]. $NbSe_2$, a superconducting dichalcogenide, has been used to study the changes in the superconductivity by field effect doping and may be used in superconducting field effect devices[5]. These materials are composed by stacks of X-M-X layers where X stands for selenium or sulfur and M for a transition metal. While atoms within the layers are strongly bonded, the layers are attracted to each other by weak van der Waals forces making these materials easy-to-cleave. This characteristic has allowed the use of the micromechanical cleavage technique to obtain atomically thin dichalcogenide crystals[1,5-7]. The micromechanical cleavage technique has been widely used to fabricate single layer graphite flakes, *i.e.* graphene, and consists on repeatedly peeling the layered bulk material and transferring this peeled material on top of a surface. This technique has been proved to be an easy and fast way of producing highly crystalline atomically thin flakes. Nevertheless micromechanical cleavage also produces a large quantity of thicker flakes making difficult the identification of the thinner ones. Scanning probe microscopies can in principle be used to discriminate the flakes by their thickness, however the low density of atomically thin flakes makes the use of such techniques unpractical. Although scanning electron microscopy is characterized by a large scanning range and fast operation, the exposition of the surfaces to the electron beam is usually accompanied by contamination of the surface. Optical microscopy[8-14] and Raman spectroscopy[15] have been used in combination with AFM to determine in a fast and non-destructive way the number of layers of graphene flakes. However it has been recently reported that Raman spectra of $NbSe_2$ do not change monotonically with the sample thickness and that ultrathin flakes can be damaged during the Raman measurements[5]. Furthermore the change on the Raman spectra with the sample thickness is still unknown for other dichalcogenide crystals. Therefore the combination of optical microscopy and AFM can be the best alternative to measure in a fast and non-destructive way the number of layers of $NbSe_2$ and $MoS_2$ thin flakes. In this context we report an experimental study of the optical contrast dependence on the $NbSe_2$ and $MoS_2$ flake thickness and the illumination wavelength. From these measurements and using a simple model we obtain the refractive index of these flakes. The study of the optical contrast dependence with the flake thickness provides a quantitative procedure to identify the number of layers of $NbSe_2$ or $MoS_2$ flakes using optical microscopy.

We have fabricated the samples by using a variation of the micromechanical cleavage used for graphene samples[16,17]. In this technique a silicone stamp[18] is used to cleave the sample instead of the commonly used Scotch tape (See supplementary material at http://dx.doi.org/10.1063/1.3442495 for more information about the technique[19]). In this way we avoid leaving traces of adhesive on the substrate which can alter the measured optical contrast[20]. Bulk $NbSe_2$ or $MoS_2$ samples are first cleaved using a viscoelastic silicone stamp. Then the cleaved flakes are transferred to an oxidized silicon wafer by pushing the stamp against the surface and pulling it slowly. Another clean stamp is used to remove the thicker flakes from the surface and eventually to recleave the thin flakes that were previously transferred.

Fig. 1(a) and 1(c) [Fig. 1(b) and 1(d)] are respectively an optical micrograph[21] and a contact mode AFM topography of an $NbSe_2$ ($MoS_2$) flake deposited onto a 90 nm $SiO_2$/Si substrate. Contact mode AFM has been chosen instead of dynamic AFM modes to avoid possible artifacts in the flake thickness measurements[22]. Fig. 1(e) and 1(f) show the height histogram of the regions marked with a dotted rectangle in Fig. 1(c) and 1(d) containing substrate, monolayer and bilayer zones. The height histograms have been fitted to three Gaussian curves to statistically determine the mean substrate, monolayer and bilayer height. The thickness of the layers and the interlayer distance can be obtained from the difference between the positions of the maxima of the peaks. In the $NbSe_2$ sample the thickness is $d = 0.9 \pm 0.1$ nm for the monolayer and $d = 1.5 \pm 0.1$ nm for the bilayer which are compatible with the interlayer distances in bulk $NbSe_2$ (0.6nm) plus the presence of a layer attributed[23] to an adsorbed water under the flake, as reported for graphene samples[23], with a thickness of 0.3 nm. In the case of $MoS_2$, the thickness for the monolayer is $d = 0.54 \pm 0.18$ nm and for the bilayer is $d = 1.1 \pm 0.2$ nm which are compatible with the bulk interlayer distances (0.6nm) if the presence of such adsorbed water layer is neglected for $MoS_2$ flakes.

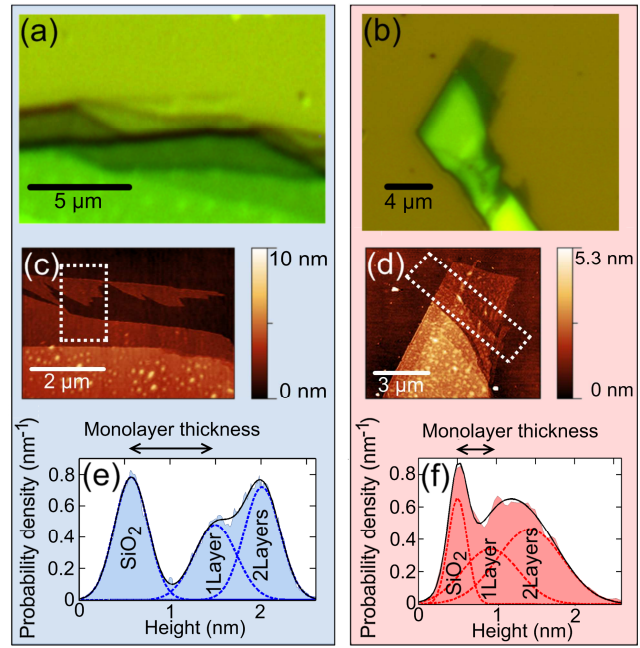

**FIG. 1.** (Color online) Optical micrographs under white light illumination of $NbSe_2$ (a) and $MoS_2$ (b) ultrathin crystals on a 90 nm $SiO_2$/Si surface. (c) and (d) are contact mode AFM topography of the same flakes in (a) and (b) respectively. Rectangles in (c) and (d) delimit zones containing the $SiO_2$ substrate, a monolayer and a bilayer. Height histogram of these zones is shown in (e) and (f). A multigaussian fit (dashed line) has been carried out to determine the layer thicknesses.

To study the optical contrast of these atomically thin flakes an approach based on the Fresnel law similar to the one developed by P. Blake *et al.*[24] has been used. The reflected intensity for normal incidence of monochromatic light can be written as[24]

$$I = \left| \frac{r_{01}e^{i(\Phi_1+\Phi_2)} + r_{12}e^{-i(\Phi_1-\Phi_2)} + r_{23}e^{-i(\Phi_1+\Phi_2)} + r_{01}r_{12}r_{23}e^{i(\Phi_1-\Phi_2)}}{e^{i(\Phi_1+\Phi_2)} + r_{01}r_{12}e^{-i(\Phi_1-\Phi_2)} + r_{01}r_{23}e^{-i(\Phi_1+\Phi_2)} + r_{12}r_{23}e^{i(\Phi_1-\Phi_2)}} \right|^2, \quad (1)$$

where the subindexes 0, 1, 2, and 3 refer to the media: air, flake, $SiO_2$ and Si respectively. $r_{ij} = (n_i - n_j)/(n_i + n_j)$ and $\Phi_i = 2\pi n_i d_i / \lambda$ with $n_i$ the complex refractive index and $d_i$ the thickness of the medium $i$. The optical contrast of the flake can be defined as

$$C = \frac{I_{flake} - I_{substrate}}{I_{flake} + I_{substraste}}. \qquad (2)$$

$I_{flake}$ can be straightforwardly determined using Eq. (1) and $I_{substrate}$ can be obtained by considering that medium 1 is air instead of a flake.

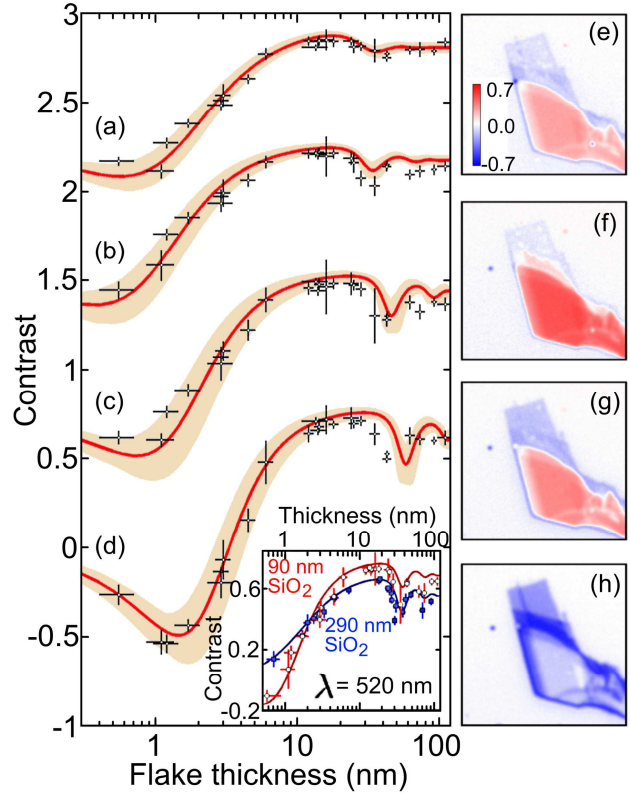

**FIG. 2.** (Color online) Measured optical contrast (symbols) of $MoS_2$ flakes onto a 90 nm $SiO_2$ substrate *vs.* their thickness at different illumination wavelengths $\lambda$: 450nm (a), 500 nm (b), 546 nm (c) and 600 nm (d). Solid red lines and shaded area are the contrast and its uncertainty obtained from the fit to Eq. (2). Note that (a), (b) and (c) have been vertically displaced for clarity by 2.25, 1.5 and 0.75 respectively. The inset below (d) shows the fit to Eq. (2) for flakes deposited on two different $SiO_2$ substrate thicknesses using the same refractive index for both cases. Contrast maps (e)-(h) obtained for the same flake in Fig. 1(b) at the same illumination wavelength as in (a)-(d). All maps share the same contrast bar [inset in (e)].

The optical contrast of $NbSe_2$ and $MoS_2$ flakes transferred on top of two different $SiO_2$/Si substrates (with 90 nm and 290 nm of $SiO_2$) have been measured as a function of their thickness at different illumination wavelengths. The thickness of the studied flakes ranges from a single layer up to 200 layers. Nine narrow band-pass filters[25] have been used to illuminate the samples at selected wavelengths in the visible spectrum. Fig. 2(a)-(d) shows the measured contrast *vs.* thickness curves for $MoS_2$ flakes transferred onto a 90 nm $SiO_2$ surface at four different illumination wavelengths. In Fig. 2(e)-(f) we show the contrast maps of the $MoS_2$ flake analyzed in Fig. 1(b). Notice that for a given flake thickness the contrast strongly depends on the illumination wavelength and interestingly thicker flakes can yield lower contrast than thinner ones. The refractive index of $NbSe_2$ and $MoS_2$ thin crystals has been obtained by fitting to Eq. (2) the experimental contrast *vs.* thickness curves measured in substrates with 90 nm or 290 nm of $SiO_2$. We have checked that a single refractive index of the flake material can be used independently of the $SiO_2$ thickness (see inset in Fig. 2) suggesting that there is no systematic error in the $SiO_2$ layer thickness. The red solid lines and the shaded area in Fig. 2(a)-(d) represent the contrast and its uncertainty obtained from the fit to Eq. (2). The agreement with the model is excellent for both $NbSe_2$ and $MoS_2$ flakes. Nevertheless we have found that the calculations systematically slightly overestimate the contrast of $NbSe_2$ flakes thinner than 2 nm. This effect for the thinnest $NbSe_2$ flakes can be attributed to a thickness dependence of the refractive index which should be more exhaustively studied.

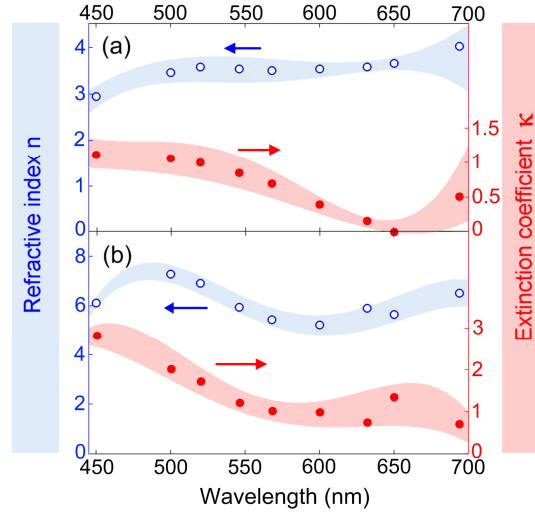

**FIG. 3.** (Color online) Determined refractive index (blue open circles, left axis) and extinction coefficient (red filled circles, right axis) for ultrathin $NbSe_2$ (a) and $MoS_2$ (b) thin crystals as a function of the incident wavelength.

The refractive index for both $NbSe_2$ and $MoS_2$ obtained at different illumination wavelengths is summarized in Fig. 3(a) and 3(b) respectively. By using these values in combination with Eq. (2) we have translated contrast maps, like the ones in Fig. 2(e)-(h), to thickness maps with uncertainties lower than 0.3 nm in thickness. Note that when the function thickness *vs.* contrast is multi-valued [see Fig. 2(c) and 2(d)] contrast maps acquired at 2 or 3 different incident wavelengths have to be used to construct the thickness maps without ambiguity. Because electronic properties of the thinnest flakes may strongly depend on the exact number of layers as in the case of graphene it is desirable a characterization method capable of unambiguously distinguish a monolayer from a bilayer, three layers… For this purpose the $SiO_2$ thickness of the substrate can be chosen to optimize the identification of $NbSe_2$ and $MoS_2$ monolayers once the refractive index of the material is known. We have calculated the $SiO_2$ thickness that simultaneously maximizes the contrast of the monolayer and the contrast difference between monolayer and bilayer. In this way for the case of $NbSe_2$ a monolayer can be better identified if a $SiO_2$ thickness of either 75 nm or 250 nm is chosen, yielding a contrast -20% for 500 nm wavelength illumination. For the case of $MoS_2$ flakes the optimal $SiO_2$ thickness is 55 nm or 220 nm and a contrast up to -60% can be obtained under an illumination wavelength of 500 nm.

In conclusion, we have fabricated and identified by optical microscopy atomically thin films of $NbSe_2$ and $MoS_2$. We have measured the dependence of the optical contrast with the flake thickness and the illumination wavelength for flakes of both materials transferred onto two different $SiO_2$/Si substrates. Interestingly the contrast is not a monotonically increasing function of the flake thickness and can even decrease as the number of layers increases. From the contrast *vs.* thickness measurements and applying a Fresnel law based model the refractive index in the visible spectrum of these thin crystals has been determined for flakes composed by more than 200 layers down to atomically thin single layers. Finally we have proposed the optimal $SiO_2$ thickness and illumination wavelength which allows for the unambiguous identification of monolayers of both materials.

**Acknowledgements**

The authors wish to thank V. Crespo and J.G. Rodrigo for providing the $NbSe_2$ bulk samples. A.C-G. acknowledges fellowship support from the Comunidad de Madrid (Spain). This work was supported by MICINN (Spain) (MAT2008-01735 and CONSOLIDER-INGENIO-2010 CSD-2007-00010) and Comunidad de Madrid (Spain) through the program Citecnomik (S_0505/ESP/0337).

# Supplemental information:

**Sample preparation:**

Scotch tape based micromechanical cleavage can be used to fabricate atomically thin dichalcogenide crystals. Nevertheless substrate contamination produced by traces of adhesive can alter the measured optical contrast[S1] and can result in an error in the determined refractive index of these atomically thin crystals. That is why it is preferable to fabricate the samples using poly (dimethyl)-siloxane (PDMS) stamps[S2], commonly known as silicone stamps, instead of adhesive tape. PDMS stamps are used in soft lithography[S3] to detach structures from a surface without the use of adhesives by pressing these structures against the stamp surface and then suddenly peeling off the stamp. Then these structures can be released on other surface by pressing the stamp surface against the new surface and peeling off the stamp very slowly. The whole process relies on the viscoelastic properties of PDMS as it behaves either like a viscous liquid or like an elastic solid depending on the peeling time-scale.

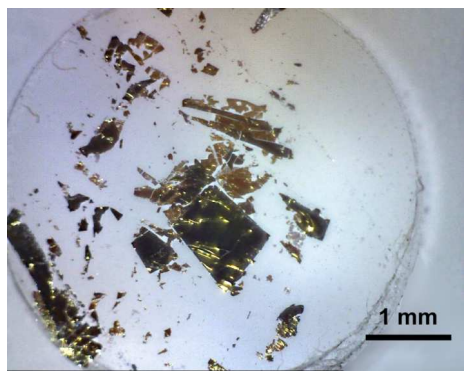

**Figure S1:** $MoS_2$ flakes attached to the PDMS stamp surface after the cleavage of a bulk $MoS_2$ crystal.

To fabricate atomically thin dichalcogenide crystals we first cleave bulk $NbSe_2$ or $MoS_2$ samples by pressing the bulk crystal surface against the stamp surface and peeling off the stamp suddenly. Figure S1 shows a detail of the PDMS stamp surface after the cleavage of a $MoS_2$ bulk crystal. Then the cleaved flakes are transferred to an oxidized silicon wafer by pressing the stamp against the surface and slowly peeling it off (~5 s to peel off completely the stamp from the surface). We use a second clean PDMS stamp to cleave again the flakes previously transferred on the $SiO_2$ surface. In this way most of the thick flakes are removed leaving only thin flakes on the surface. We have found that thin flakes are eventually re-cleaved resulting in thinner flakes. A closer inspection of the samples reveals that the flakes edges are very straight and tend to form well defined angles (figure S2) as expected for crystalline samples. This is also characteristic of graphene samples obtained by the micromechanical cleavage technique using Scotch tape or PDSM stamps[S2].

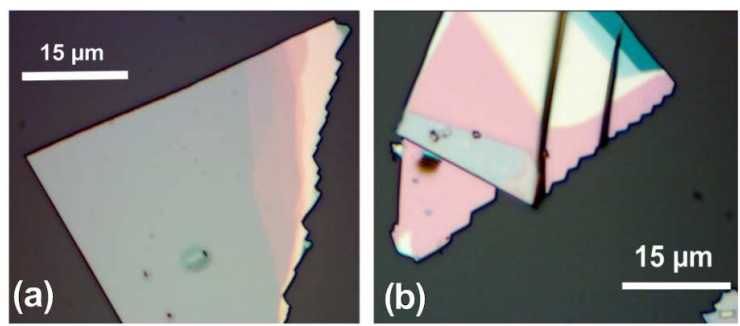

**Figure S2:** Multilayered $NbSe_2$ flake (a) and $MoS_2$ (b) deposited on a 90 nm $SiO_2$/Si surface.

**Silicone stamps casting:**

The procedure to fabricate the silicone stamps in a reproducible way is the following:

We use a Teflon mold to cast the silicone stamps into an easy to handle shape (figure S3a).

Press the mold against a clean and flat surface like a glass slide or a silicon chip. Note that the part of the stamp used to cleave the crystals is the one in contact with this flat and clean surface (working surface).

Mix the 184 Sylgard polymer base and the curing agent in a 10:1 ratio by weight.

Pour the mixed 184 Sylgard into the mold.

Place the filled mold in a low vacuum chamber (500 mbar during 1 hour) to degas the mixed 184 Sylgard. In this way air bubbles are removed and the working surface of the stamp will be flat.

Place the filled mold in an oven at 60 ºC during 24 hours.

Unmold the stamp. A mold of Teflon (figure S3a) facilitates the unmolding process and prevents from breaking the stamp.

The resulting stamps are shown in figure S3b.

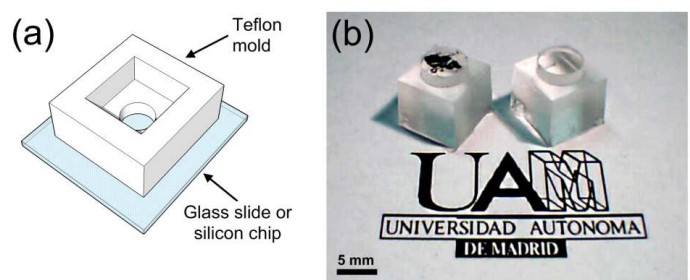

**Figure S3.** (a) Schematic of the mold/surface assembly used in the casting procedure. (b) Picture of two PDMS stamps fabricated using the described recipe. The one on the left shows NbSe$_2$ flakes attached to its working surface.